%Paper: hep-th/9410225
%From: YVONNE@urhep.pas.rochester.edu
%Date: Fri, 28 Oct 1994 13:37:24 -0500 (EST)
%Date (revised): Fri, 28 Oct 1994 14:18:24 -0500 (EST)

\magnification=1200
\baselineskip=18pt
\tolerance=100000
\overfullrule=0pt

\centerline{\bf RELATIVISTIC AHARONOV-BOHM-COULOMB PROBLEM}

\vskip 1truein

\centerline{C. R. Hagen and D. K. Park\footnote*{On leave from Dept. of
Physics, Kyung Nam University, Masan, 631-701, Korea}}
\centerline{Department of Physics and Astronomy}
\centerline{University of Rochester}
\centerline{Rochester, NY 14627, USA}

\vskip 2 truein

\noindent {\bf Abstract}

\medskip

The Aharonov-Bohm effect is analyzed for a spin-1/2 particle in the case
that a $1/r$ potential is present.  Scalar and vector couplings are each
considered.  It is found that the approach in which the flux tube is
given a finite radius that is taken to zero only after a matching of
boundary conditions does not give physically meaningful results.
Specifically, the operations of taking the limit of zero flux tube radius
and the Galilean limit do not commute.  Thus there appears to be no
satisfactory solution of the relativistic Aharonov-Bohm-Coulomb problem
using the finite radius flux tube method.

\vfill\eject

\noindent {\bf 1. Introduction}

\medskip

Since the time of its discovery the Aharonov-Bohm (AB) effect [1] has
been the object of considerable attention at both theoretical and
experimental levels [2].  Much of the recent attention given to this
phenomenon has been
 associated with the fact that
 for spinless nonrelativistic particles
  it allows an interpretation
in which the interaction can be eliminated
provided that  fractional statistics are
 introduced [3].  In fact it has been suggested that
high-$T_c$ superconductivity phenomena may be best understood using
fractional statistics.  That interpretation, however, cannot be
maintained when one considers spin effects in the AB problem, since the
latter has the property of introducing a
 delta-function potential into the Hamiltonian (i.e.,
the Zeeman interaction of the spin with the magnetic field).  This term
breaks an essential symmetry in the fractional statistics approach,
namely the invariance of the theory under translation of the flux
parameter by an integer.  The spin-${1 \over 2}$ AB problem has been
discussed extensively by different methods.  Gerbert [4] examined the
problem from a mathematical point of view by the self-adjoint extension
approach and concluded that an arbitrary combination of the regular and
singular solutions could contribute to the wave function provided that
$$|m + \alpha | < 1 \quad , \eqno(1.1)$$
where $\alpha$ is the flux parameter.  The same problem was subsequently
considered by one of us [5] in  the framework of a  more physical model.
 Specifically, the  magnetic field in the  Zeeman
interaction term was defined in that approach to be
 the zero radius  limit of a flux tube of finite radius, i.e.,
 is proportional to
$$\lim_{R \rightarrow 0} {\alpha \over R} \ \delta  (r -R)
\quad . \eqno(1.2)$$
\noindent From the boundary conditions at $r=R$
 it was then found that only the
singular solution of the Schr\"odinger equation
  could contribute to the radial wave equation for the case
$$\eqalign{&|m| + |m + \alpha | = - \alpha s\cr
&|m+ \alpha | < 1\cr}\eqno(1.3)$$
where $s$ is twice the spin projection.

In order to achieve a physical realization of the self-adjoint extension
method some authors [6] have recently considered the possibility of including
strongly repulsive potentials inside the flux tube. However, it has been
noted [7] that such calculations within the framework of the Dirac
equation tend not to be reliable because of the occurrence of Klein's
paradox.  To avoid this difficulty the solution of the spin-${1 \over 2}$
 Aharonov-Bohm-Coulomb (ABC) problem [7]
 was obtained within the framework of  the
 Galilean theory [8] which is free
 of such complications.  It was then found that
 with the inclusion of the Coulomb
 potential,  the range of flux parameter for which singular
 solutions are allowed is
 only half as large as that in the pure AB case, namely,
 $$| m + \alpha | < {1 \over 2} \quad . \eqno(1.4)$$
 Subsequently it was shown [9] that the self-adjoint extension method
 also gives the condition (1.4) for the occurrence of singular solutions.

 The goal of this paper is to carry out an analysis of the ABC problem
 without recourse to the Galilean limit.
  It should be noted at the outset, however, that there
 are at least two ways in which a Coulomb potential
  can be included in  the
 relativistic AB problem
 such that
 the same Galilean limit is obtained.
 Thus the analysis presented here considers both
  the scalar coupling defined by
 $$M \rightarrow M + \xi / r \eqno(1.5)$$
 (where $M$ is the mass of the spin-${1 \over 2}$ particle and $\xi/r$ is
 the Coulomb potential)
  as well as the vector coupling
 $$E \rightarrow E - \xi /r \quad . \eqno(1.6)$$
   Furthermore,
   two different realizations of the Coulomb potential within the
 flux tube are considered in this paper,
  one continuous $(V_c)$ and one discontinuous
 $(V_d)$.  They are parametrized as
 $$\eqalignno{V_c (r) &= {\xi \over r}\ \theta (r-R) + {\xi \over
 R}\ \theta (R-r)&(1.7a)\cr
 \noalign{\vskip 4pt}%
 V_d (r) &= {\xi \over r}\ \theta (r-R) &(1.7b)\cr}$$
 where $\theta (x)$ is the usual step function
 $$\theta (x) = {1 \over 2}\ \left( 1 + {x \over |x|} \right) \quad . $$
 In secs. 2 and 3 the Dirac equation of the ABC problem with a scalar
 coupling is solved explicitly,  first for
 the case of the continuous
  potential (1.7a) and  subsequently for the  discontinuous one (1.7b).
   The comparisons between the relativistic
 AB and Galilean ABC problems are discussed in detail.  Corresponding
  results for the
 vector coupling of the Coulomb potential
  are presented in sec. 4.  A concluding section compares results
   in the Galilean and relativistic ABC problems and makes some
  general observations concerning the significance of the results obtained.

 \medskip

 \noindent {\bf 2. Scalar Coupling of the Continuous Coulomb
 Potential}

 \medskip

 In this section the relativistic ABC problem is analyzed for the case in
 which the Coulomb potential (1.7a) is included in the Dirac equation by
 means of a scalar interaction.  Thus the relevant equation is
 $$\beta \{[ M+V(r) ] + \vec \gamma \cdot
 \vec \Pi \} \psi = E \psi \eqno(2.1)$$
where $\Pi_i = - i \partial_i - eA_i$ with $A_i$ the usual AB potential
$$e A_i = \cases{{\alpha \epsilon_{ij} r_j \over r^2}  & $r > R$\cr
 \noalign{\vskip 6pt}%
0  &$r<R$ \quad .\cr} \eqno(2.2)$$
A convenient choice for the matrices in (2.1) is [5]
$$\eqalign{&\beta = \sigma_3\cr
&\beta \gamma_i = (\sigma_1 , s \sigma_2 )\cr}\eqno(2.3)$$
where the $\sigma$'s are the usual Pauli matrices and $s = \pm 1$ is the spin
projection parameter of Eq. (1.3).

The second order equation implied by (2.1) is obtained by applying the
matrix operator $[(M + V(r)) + \beta E - \vec \gamma \cdot
\vec \Pi ] \beta$.  The result is
$$\eqalign{\bigg\{ {\partial^2 \over \partial r^2} + {1 \over r}
\ {\partial
\over \partial r} &+ { 1 \over r^2}\ \bigg( {\partial \over
\partial \phi}
+ i \alpha\bigg)^2 \cr
&+ E^2 - [M + V(r)]^2 +
esH \sigma_3 + E_1 \sigma_2 - s E_2 \sigma_1 \bigg\} \psi =0 \cr}
\eqno(2.4)$$
where $H$ is the magnetic field
$$esH = - {\alpha s \over R}\ \delta(r-R) \eqno(2.5)$$
and $E_i$ is the electric field
$$E_i = - \partial_i V(r) \quad .\eqno(2.6)$$
Using (1.7a) one finds that
$$E_1 \pm is E_2 = {\xi \over r^2} \ e^{\pm is \phi} \theta (r-R)
\eqno(2.7)$$
 so that (2.4) becomes
$$\eqalign{\bigg\{{\partial^2 \over \partial r^2} &+ {1 \over r}\ {\partial
\over \partial r} + {1 \over r^2}\ \big( {\partial \over \partial \phi}
+ i\alpha \big)^2 + E^2
- \bigg[ M + {\xi \over r} \theta\  (r-R) +
{\xi \over R}\ \theta (R-r)\bigg]^2\cr
  &+ es H \sigma_3 \bigg\}
\psi = - {\xi \over r^2}\ \pmatrix{0 & -ie^{-is\phi}\cr
 \noalign{\vskip 4pt}%
ie^{is\phi} &0 \cr} \theta (r-R) \psi \quad .\cr}
 \eqno(2.8)$$
If one defines
$$\psi = \pmatrix{\psi_1\cr
 \noalign{\vskip 5pt}%
\psi_2\cr} = \pmatrix{\sum_m f_m (r) e^{im \phi}\cr
 \noalign{\vskip 5pt}%
\sum_m g_m (r) e^{i (m+s)\phi}\cr} \quad , \eqno(2.9)$$
Eq. (2.8) becomes
$$\eqalign{\bigg\{ {d^2 \over dr^2} &+ {1 \over r}\ {d
\over dr} - {(m + \alpha )^2 \over r^2} + E^2
 - \bigg[ M + {\xi \over r}\  \theta (r-R) +
{\xi \over R}\ \theta (R-r)\bigg]^2 \cr
&- { {1 \over 2}\ +s (m+\alpha) \over r^2}\
(1 - \sigma_3)
+ {\xi \over r^2}\ \sigma_2 \theta (r-R) +
  esH \sigma_3 \bigg\}
  \pmatrix{f_m\cr
 \noalign{\vskip 5pt}%
  g_m\cr} = 0 \quad . \cr}\eqno(2.10)$$

\indent From (2.10) the equation governing the inside region $(r < R)$ is seen
to be
$$\left[ {d^2 \over dr^2} + {1 \over r}\ {d \over dr} - {m^2 \over r^2} +
E^2 - \left( M + {\xi \over R}\right)^2 \right] f_m = 0\eqno(2.11)$$
whose regular solution is
$$f^{in}_m (r) = J_{|m|} (k_0 r) \eqno(2.12)$$
where
$$k^2_0 = E^2 - \left( M + {\xi \over R}\right)^2 \quad .\eqno(2.13)$$
In the outside region $(r > R)$ Eq. (2.10) becomes
$$\left[ {d^2 \over dr^2} + {1 \over r}\ {d \over dr} -
{\eta + (m + \alpha)^2 \over r^2} + E^2 -
\left( M + {\xi \over r}\right)^2 + {1 \over r^2} \left( \eta \sigma_3
+ \xi \sigma_2 \right) \right]
\pmatrix{f_m \cr
g_m\cr} = 0 \eqno(2.14)$$
where
$$\eta = s(m + \alpha ) + {1 \over 2} \quad .\eqno(2.15)$$
It is convenient to bring the matrix $\eta \sigma_3 + \xi \sigma_2$ to
 diagonal form using
$$\eta \sigma_3 + \xi \sigma_2 = \epsilon (\eta) \sqrt{\eta^2 + \xi^2}\
U^+ \ \sigma_3 \ U \eqno(2.16)$$
where $\epsilon (x)$ is the usual alternating function $\epsilon
(x) = x /|x|$ and the unitary matrix $U$ is
$$U = \cos {\theta \over 2} - i \sigma_1 \epsilon (\eta)
\sin {\theta \over 2}$$
\line{with \hfill (2.17)}
$$- \pi  < \theta = \tan^{-1} {\xi \over \eta} < \pi \quad .$$

By defining
$$\chi_{_m} = \pmatrix{\chi_{_{1m}}\cr
 \noalign{\vskip 5pt}%
\chi_{_{2m}}\cr} = U \pmatrix{f_m\cr
 \noalign{\vskip 5pt}%
g_m\cr} \quad , \eqno(2.18)$$
the second order equation for $\chi_m$ becomes
$$\eqalign{\bigg[ {d^2 \over dr^2} &+ {1 \over r}\ {d \over dr} -
{(m +\alpha)^2 + \eta  \over r^2} +
E^2\cr
&- \bigg( M + {\xi \over R}\bigg)^2
+ {\epsilon (\eta) \sqrt{\eta^2 +\xi^2} \over
r^2} \ \sigma_3 \bigg]
 \chi_{_m} = 0 \quad . \cr}\eqno(2.19)$$
The solutions of Eq. (2.19) are
$$\eqalign{\chi_{_{1m}} &= A_m e^{ikr} (-2 ikr)^{a_m} F \bigg( a_m + {1
\over 2} + {i
M\xi \over k} \ | 2a_m +1 | -2ikr \bigg)\cr
 \noalign{\vskip 4pt}%
&\qquad + B_m e^{ikr} (-2ikr)^{-a_m} F \bigg( -a_m + {1 \over 2} +
{iM\xi \over k} \ |1- 2a_m | - 2ikr \bigg)\cr
 \noalign{\vskip 4pt}%
\chi_{_{2m}} &= C_m e^{ikr} (-2 ikr)^{b_m} F \bigg( b_m + {1
\over 2} + {i
M\xi \over k} \ | 2b_m +1 | -2ikr \bigg)\cr
 \noalign{\vskip 4pt}%
&\qquad + D_m e^{ikr} (-2ikr)^{-b_m} F \bigg( -b_m + {1 \over 2} +
{iM\xi \over k} \ |1- 2b_m | - 2ikr \bigg)\cr}\eqno(2.20)$$
where $F(a| c|z)$ is the usual confluent hypergeometric function and
$$\eqalign{k^2 &= E^2 - M^2\cr
 \noalign{\vskip 4pt}%
a_m &= \sqrt{\eta^2 + \xi^2} - {\epsilon (\eta) \over
2}\cr
 \noalign{\vskip 4pt}%
b_m &= \sqrt{\eta^2 + \xi^2} + {\epsilon (\eta) \over
2} \quad . \cr}\eqno(2.21)$$
If one inverts Eq. (2.18), $f_m$ and $g_m$ are inferred to be
$$\eqalign{f^{\rm out}_m (r) &= \cos {\theta \over 2}\ \chi_{_{1m}}
+ i \epsilon (\eta) \sin {\theta \over 2}\ \chi_{_{2m}}\cr
 \noalign{\vskip 4pt}%
g^{\rm out}_m (r) &= i \epsilon (\eta)
\sin {\theta \over 2}\ \chi_{_{1m}}
+  \cos {\theta \over 2}\ \chi_{_{2m}}\quad . \cr}\eqno(2.22)$$
Before proceeding it is worthwhile to note the Galilean limit of Eq.
(2.22).  Since in that regime
$$\eqalign{&\cos \theta \rightarrow 1\cr
&\sin \theta \rightarrow 0\cr
&a_m \rightarrow |\eta| - {\epsilon (\eta) \over 2}\cr
 \noalign{\vskip 4pt}%
&b_m \rightarrow |\eta| + {\epsilon (\eta) \over 2}\quad , \cr}
\eqno(2.23)$$
the Galilean limit of $f^{\rm out}_m (r)$ is
$$\eqalign{f^{\rm out}_m (r) &\rightarrow A_m e^{ikr} (-2i kr)^{|\eta| -
{\epsilon (\eta) \over 2}}\cr
& F \bigg( |\eta | - {\epsilon (\eta) \over 2} +
{1 \over 2} + {i M \xi \over k} \ \bigg| 2 (|\eta | -
 {\epsilon (\eta) \over 2} \bigg) + 1 \bigg| -2ikr\bigg)\cr
& + B_m e^{ikr} (-2i kr)^{-|\eta| +
{\epsilon (\eta) \over 2}}\cr
& F \bigg( -|\eta | + {\epsilon (\eta )
 \over 2} +
{1 \over 2} + {i M \xi \over k} \ \bigg| 1-2 \bigg( |\eta | -
 {\epsilon (\eta) \over 2} \bigg) \bigg| -2ikr\bigg)\ \ .\cr}\eqno(2.24)$$
  By considering the four cases

  \item{(1)} $s = 1  \qquad - {1 \over 2} \ < m+\alpha < 1$
  \item{(2)} $s = -1  \qquad -1 < m + \alpha < {1 \over 2}$
  \item{(3)} $s= 1 \qquad -1 < m + \alpha < - {1 \over 2}$
  \item{(4)} $s=-1 \qquad {1 \over 2} < m + \alpha < 1 \qquad$

 \noindent respectively, one can show that
 for $| m + \alpha | < 1$ Eq. (2.24) becomes
 $$\eqalign{A_m &e^{ikr} (-2i kr)^{|m + \alpha |} F \bigg( |m+\alpha | +
 {1 \over 2} + {iM\xi \over k} \ \big| 1+ 2 |m+\alpha | \big| - 2ikr
 \bigg)\cr
  &+ B_m e^{ikr} (-2i kr)^{-|m + \alpha |} F \bigg( -|m+\alpha | +
 {1 \over 2} + {iM\xi \over k} \ \big| 1- 2 |m+\alpha | \big| - 2ikr
 \bigg)\cr}$$
 as already derived in Ref. [7].

 Before  one considers the boundary
 condition at $r=R$, it is important to derive the relations between  the
 coefficients $A_m$, $B_m$, $C_m$ and $D_m$ by using the first-order
 Dirac equation (2.1).  From (2.1) the first-order equations for $f_m
 (r)$ and $g_m (r)$ are
 $$\eqalign{
 \bigg( M - {\xi \over r} - E \bigg) f_m &= i \bigg( {d \over dr} + {s
 (m + \alpha + s) \over r} \bigg) g_m\cr
 \noalign{\vskip 4pt}%
 -\bigg( M - {\xi \over r} + E \bigg) g_m &= i \bigg( {d \over dr} - {s
 (m + \alpha ) \over r} \bigg) f_m\cr}\eqno(2.25)$$
 which yield the coupled first order equations for $\chi_{_{1m}}$ and
 $\chi_{_{2m}}$
 $$\eqalign{{d \chi_{_{1m}} \over dr} &+ \bigg[ {\epsilon (\eta) \xi
 \over \sqrt{\eta^2 + \xi^2}} \ M - {\epsilon (\eta)
 \sqrt{\eta^2 + \xi^2} - {1 \over 2} \over r} \bigg] \chi_{_{1m}}\cr
 & = i
 \bigg( E +M {|\eta | \over \sqrt{\eta^2 + \xi^2}} \bigg) \chi_{_{2m}}\cr
 \noalign{\vskip 4pt}%
 {d \chi_{_{2m}} \over dr} &- \bigg[ {\epsilon (\eta) \xi
 \over \sqrt{\eta^2 + \xi^2}} \ M - {\epsilon (\eta)
 \sqrt{\eta^2 + \xi^2} +{1 \over 2} \over r} \bigg] \chi_{_{2m}}\cr
 & = i
 \bigg( E -M {|\eta | \over \sqrt{\eta^2 + \xi^2}} \bigg) \chi_{_{1m}}
 \quad . \cr}
 \eqno(2.26)$$
 By inserting $\chi_{_{1m}}$ and $\chi_{_{2m}}$ into Eq. (2.26) the relations
 between coefficients are seen to be
 $$\eqalign{C_m &= \Omega_1 A_m\cr
 D_m &= \Omega_2 B_m\cr}$$
 \line{when $\epsilon (\eta ) = 1$, and \hfill (2.27)}
   $$\eqalign{A_m &= \Omega_1 C_m\cr
 B_m &= \Omega_2 D_m\cr}$$
when $\epsilon (\eta ) = -1$ where
$$\eqalign{\Omega_1 &= {-k \big( \eta^2 + \xi^2 + M^2 \xi^2
 k^{-2}\big) \over
2 \big( \eta^2 + \xi^2 \big) \big( 2 \sqrt{\eta^2 + \xi^2} + 1 \big)} \
\bigg( E + {\eta \over \sqrt{\eta^2 + \xi^2}}\ M \bigg)^{-1}\cr
 \noalign{\vskip 4pt}%
\Omega_2 &= 2 k \big( 2 \sqrt{\eta^2 + \xi^2} - 1 \big) \bigg(
E + {\eta \over \sqrt{\eta^2 + \xi^2}} \ M \bigg)^{-1} \quad
. \cr} \eqno(2.28)$$
It can be shown by direct calculation that the $\xi
\rightarrow 0$ limit of Eq.
(2.27) is identical to the result
 obtained if one starts {\it ab initio}  from the
relativistic AB problem.

 By using Eq. (2.27) the outside solution of
$f_m (r)$ can be expressed as
$$f^{\rm out}_m (r) = \cases{A_m u_1 (r) + B_m u_2 (r) &when
$\quad \epsilon (\eta) = 1$\cr
 \noalign{\vskip 6pt}%
C_m v_1 (r) + D_m v_2 (r) &when
$\quad  \epsilon (\eta) = -1$\cr}\eqno(2.29)$$
where
$$\eqalign{u_1 (r) = e^{ikr} \bigg[
&\cos {\theta \over 2} \
(-2ikr)^{\beta_{m,s}} F \bigg( \beta_{m,s}
 + {1 \over 2} + {iM \xi \over k}
\big| 2 \beta_{m,s} + 1 \big| -2ikr\bigg)\cr
 \noalign{\vskip 6pt}%
&+ i \sin {\theta \over 2} \ \Omega_1 (-2ikr)^{\beta_{m,s} +1}\ F
\bigg( \beta_{m,s} + {3 \over 2} + {iM\xi \over k} \big| 2 \beta_{m,s} + 3
\big| -2i kr \bigg)\bigg]\cr
 \noalign{\vskip 4pt}%
 u_2 (r) = e^{ikr} \bigg[&\cos {\theta \over 2} \
(-2ikr)^{-\beta_{m,s}} F \bigg( -\beta_{m,s}
 + {1 \over 2} + {iM \xi \over k}
\big| 1-2 \beta_{m,s} \big|  -2ikr\bigg)\cr
 \noalign{\vskip 4pt}%
&+ i \sin {\theta \over 2} \ \Omega_2 (-2ikr)^{-\beta_{m,s} -1} F
\bigg( -\beta_{m,s} - {1 \over 2} + {iM\xi \over k} \big| -1 -2
 \beta_{m,s}
\big| -2i kr \bigg)\bigg]\cr
 \noalign{\vskip 4pt}%
 v_1 (r) = e^{ikr} \bigg[&\cos {\theta \over 2} \
\Omega_1 (-2ikr)^{\beta_{m,s}+1} F \bigg( \beta_{m,s}
 + {3 \over 2} + {iM \xi \over k}
\big| 2 \beta_{m,s} + 3 \big| -2ikr\bigg)\cr
 \noalign{\vskip 6pt}%
&+ i \sin {\theta \over 2} \  (-2ikr)^{\beta_{m,s}} F
\bigg( \beta_{m,s} + {1 \over 2} + {iM\xi \over k} \big| 2 \beta_{m,s} + 1
\big| -2i kr \bigg)\bigg]\cr
 \noalign{\vskip 4pt}%
v_2 (r) = e^{ikr} \bigg[&\cos {\theta \over 2} \ \Omega_2
(-2ikr)^{\beta_{m,s}-1} F \bigg( -\beta_{m,s}
 - {1 \over 2} + {iM \xi \over k}
\big| -2 \beta_{m,s} - 1 \big| -2ikr\bigg)\cr
 \noalign{\vskip 6pt}%
&- i \sin {\theta \over 2} \  (-2ikr)^{-\beta_{m,s}} F
\bigg( -\beta_{m,s} + {1 \over 2} + {iM\xi \over k} \big| 1- 2 \beta_{m,s}
\big| -2i kr \bigg)\bigg]\cr}\eqno(2.30)$$
and
$$\beta_{m,s} = \sqrt{\eta^2 + \xi^2} - {1 \over 2}\quad . \eqno(2.31)$$
The boundary conditions at $r=R$
$$\eqalign{&\lim_{\varepsilon \rightarrow 0^+} f_m (R+\varepsilon)
= \lim_{\varepsilon \rightarrow 0^+} f_m (R-\varepsilon)\cr
 \noalign{\vskip 4pt}%
&\lim_{\varepsilon \rightarrow 0^+} {df_m  \over dr}\ (R+\varepsilon)
= \lim_{\varepsilon \rightarrow 0^+} \bigg( {d \over dr}
+ {\alpha s \over R} \bigg) f_m (R-\varepsilon)\cr}\eqno(2.32)$$
give the ratios of the coefficients as
$$\eqalign{{A_m \over B_m} &= {J_{|m|} (k_0 R) {du_2 (R) \over dR} - u_2 (R)
\big( {d \over dR} + {\alpha s \over R}\big) J_{|m|} (k_0 R) \over
u_1 (R) \big( {d \over dR} + {\alpha s \over R} \big) J_{|m|} (k_0 R) -
J_{|m|} (k_0 R) \ {d u_1 (R) \over dR}}\cr
 \noalign{\vskip 4pt}%
{C_m \over D_m} &= {J_{|m|} (k_0 R) {dv_2 (R) \over dR} - v_2 (R)
\big( {d \over dR} + {\alpha s \over R}\big) J_{|m|} (k_0 R) \over
v_1 (R) \big( {d \over dR} + {\alpha s \over R} \big) J_{|m|} (k_0 R) -
J_{|m|} (k_0 R) \ {d v_2 (R) \over dR}}
\cr}\eqno(2.33)$$
for $\epsilon (\eta) = 1$ and $\epsilon (\eta ) = -1$ respectively,
where ${d \over dR}\ J_{|m|} (k_0R)$ is understood to be $\left[ {d \over
dr}\ J_{|m|} (k_0 r) \right]_{r=R}$.  At this point it is appropriate to
take the $R \rightarrow 0$ limit.  In order to compare with various other
AB problems the first and second leading terms of $u_1(r)$, $u_2 (r)$,
$v_1 (r)$ and $v_2 (r)$ are given in Tables I, II, III, and IV
respectively.  From these tables it is seen that the Galilean limit of
the exponents of $R$ for the first leading terms of $u_2 (R)$ and $v_1
(R)$ in the relativistic ABC problem do not coincide with those of the
Galilean ABC problem. In other words, the Galilean limit operation does
not commute with the  $R \rightarrow 0$ limit.  This feature
makes the relativistic ABC problem quite different from the Galilean
one.  Another factor which exacerbates this difference is the
$R$-dependence of $k_0$.  While in the relativistic ABC problem $k_0$ is
proportional to $R^{-1}$ for $R \rightarrow 0$, in the Galilean limit
$k_0$ is proportional to $R^{- {1 \over 2}}$.  The first and
second leading terms of $J_{|m|} (k_0 R)$ and $\left( {d \over dR} +
{\alpha s \over R}\right) J_{|m|} (k_0 R)$ in the various AB problems are
given in Table V and VI.

If one applies the first and second leading
terms of the relativistic AB and Galilean ABC problems to Eq. (2.33), the
conditions for the occurrence of a singular solution are
$$\eqalign{&|m+ \alpha | < 1\cr
&|m| + |m + \alpha | + \alpha s = 0\cr}\eqno(2.34)$$
 in the relativistic AB problem and
$$\eqalign{&|m+ \alpha | < {1 \over 2}\cr
&|m| + |m + \alpha | + \alpha s = 0  \cr}\eqno(2.35)$$
for the Galilean ABC case.
However, the nature of the relativistic ABC problem is vastly different.
Of particular significance is the fact that for nonzero values of
$\theta$ the functions $u_1$, $u_2$, $v_1$, and $v_2$ can each have
both  regular \underbar{and} irregular parts.  By applying Eq. (2.33)
one concludes that
$$B_m = D_m = 0 \eqno(2.36)$$
so long as there is no fine tuning of $\xi$.  Thus the radial solution
becomes
$$\eqalign{f_m(r) =
A_m e^{ikr} \bigg[&\cos {\theta \over 2}\
(-2ikr)^{\beta_{m,s}} F \bigg( \beta_{m,s} + {1 \over 2} +
{iM\xi \over k} | 2\beta_{m,s} + 1 | -2ikr\bigg)\cr
&+ i \sin {\theta \over 2}\ \Omega_1 (-2i kr)^{\beta_{m,s}+1}
F\bigg( \beta_{m,s} + {3 \over 2} + {iM\xi \over k} |
2 \beta_{m,s} + 3 | -2ikr \bigg)\bigg]\cr}\eqno(2.37)$$
for $\epsilon (\eta ) = 1$, and
$$\eqalign{f_m (r) = C_m e^{ikr} \bigg[&\cos {\theta \over 2}\  \Omega_1
(-2ikr)^{\beta_{m,s}+1} F \bigg( \beta_{m,s} + {3 \over 2} +
{iM\xi \over k} | 2\beta_{m,s} + 3 | -2ikr\bigg)\cr
&+ i \sin {\theta \over 2}\  (-2i kr)^{\beta_{m,s}}
F\bigg( \beta_{m,s} + {1 \over 2} + {iM\xi \over k} |
2 \beta_{m,s} + 1 | -2ikr \bigg)\bigg]\cr}\eqno(2.38)$$
for $\epsilon (\eta) = -1$.
\noindent From (2.36)  bound state energies are obtained
 for the $\xi < 0$ case by applying
simultaneously the terminating condition to the two confluent
hypergeometric functions.  The result is
$$k \equiv \sqrt{E^2 -M^2} =
-{iM\xi \over n-1 +  [ \eta^2 + \xi^2]^{1/2}}
\qquad\qquad n = 1,2,\dots
\eqno(2.39)$$
where use has been made of the fact that $\Omega_1$ vanishes in the
$n=1$ state.  Detailed discussion of this expression for the
binding energies is deferred to the Conclusion.

\medskip

\noindent {\bf 3. Scalar Coupling of the Discontinuous Coulomb
Potential}

\medskip

The most distinctive feature of the discontinuous Coulomb potential
relative to the continuous one is that the electric field acquires a
delta-function contribution which changes the boundary condition at
$r=R$.  Thus Eq. (2.7) in this case becomes
$$E_1 \pm is E_2 = e^{\pm is\phi} \left[ {\xi \over r^2}\ \theta (r-R) -
{\xi \over R}\ \delta (r-R) \right]  \eqno(3.1)$$
while the boundary conditions at $r=R$ are
$$\eqalign{&\lim_{\varepsilon \rightarrow 0^+} f_m (R +
\varepsilon) =
\lim_{\varepsilon \rightarrow 0^+} f_m (R-\varepsilon)\cr
&\lim_{\varepsilon \rightarrow 0^+} g_m (R +\varepsilon) =
\lim_{\varepsilon \rightarrow 0^+} g_m (R-\varepsilon)\cr
&\lim_{\varepsilon \rightarrow 0^+} \ {df_m \over dr} \ (R+\varepsilon) =
\lim_{\varepsilon \rightarrow 0^+} \bigg( {d \over dr} +{\alpha s \over
R}\bigg) f_m (R -\varepsilon ) - {i \xi \over R}\ g_m (R) \quad .\cr}
\eqno(3.2)$$
Since $g_m$ is included in the boundary conditions, one must solve for
both $f_m (r)$ and $g_m (r)$ in the inside region.  One readily
establishes from the radial equation  for the inside
 region that
 $$\eqalign{f^{\rm in}_m (r) &= E_m J_{|m|} (kr)\cr
 g^{\rm in}_m (r) &= F_m J_{|m+s|} (kr)\cr}\eqno(3.3)$$
 where
 $$k^2 = E^2 -M^2\quad . \eqno(3.4)$$
 By using a first-order equation
  it is found that the relation between $E_m$ and $F_m$
  can be written as
 $$F_m = i \epsilon (ms) \ \sqrt{{E-M \over E+M}} \ E_m
 \eqno(3.5)$$
 provided that $\epsilon (0)$ is taken to be $+1$.

It is easily demonstrated that the outside solution is given by Eq.
(2.29), just as in the case of the continuous Coulomb potential.  By
using the inside solutions (3.3) and the outside solutions (2.29) the
boundary conditions (3.2) give the ratio of coefficients
$${A_m \over B_m} = {J_{|m|} (kR) \ {du_2 (R) \over dR} - u_2
(R) u(R) \over u_1 (R) u(R) - J_{|m|} (kR) \ {du_1 (R) \over dR}}
\eqno(3.6)$$
for $\epsilon (\eta) = 1$, and
$${C_m \over D_m} = {J_{|m|} (kR) \ {dv_2 (R) \over dR} - v_2
(R) u(R) \over v_1 (R) u(R) - J_{|m|} (kR) \ {dv_1 (R) \over dR}}
 \eqno(3.7)$$
 for $\epsilon (\eta) = -1$ where
$$u(R) = \left( {d \over dR} + {\alpha s \over R}\right) J_{|m|} (kR) +
{\xi \over R}\ \sqrt{{E-M \over E+M}} \ \epsilon (ms) J_{|m+s|} (kR)
\quad . \eqno(3.8)$$
In order to compare with the various AB problems the first and second
leading terms of $u(R)$ in the $R \rightarrow 0$ limit are given in Table
VII.  It is important to note that if one substitutes these asymptotic
forms for $u(R)$ into Eqs. (3.6)
 and (3.7) for the relativistic AB and Galilean ABC
systems, Eq. (2.34) and (2.35) are again derived
 for the occurrence of  singular
solutions.  In the relativistic ABC problem one again obtains  Eqs. (2.37)
 and (2.38) if  fine tuning of $\xi$ is not considered.

\medskip

\noindent {\bf 4. Vector Coupling Theory }

\medskip

In this section the results which have been derived
in the preceding sections are extended to
the vector coupling of a Coulomb potential in the relativistic AB
problem.  The Dirac equation to be discussed is
$$\beta \left[ M + \vec \gamma \cdot \vec \Pi \right] \psi =
[E - V(r)] \psi
\eqno(4.1)$$
which implies the second-order equation
$$\eqalign{\bigg[ {\partial^2 \over \partial r^2} &+ {1 \over r} \
{\partial \over \partial r} + {1 \over r^2}\ \bigg( {\partial \over
\partial \phi} + i \alpha \bigg)^2\cr
&+ esH \sigma_3 + (E -V)^2 - M^2 -i (E_1 \sigma_1 + s E_2 \sigma_2) \bigg]
\psi = 0 \quad . \cr}\eqno(4.2)$$
By using Eq. (2.9) the radial equation for the continuous Coulomb
potential is seen to be
$$\eqalign{\bigg\{{d^2 \over dr^2} &+ {1 \over r}\ {d \over dr} - {(m+\alpha
)^2 \over r^2} + \bigg[ E- {\xi \over r}\ \theta (r-R) - {\xi \over R}\
\theta (R-r)\bigg]^2 -M^2\cr
&-{{1 \over 2} - s(m+\alpha)\over r^2} \ (1 - \sigma_3 ) - {i \xi \over
r^2}\ \sigma_1 \theta (r-R) + esH \sigma_3 \bigg\}
 \pmatrix{f_m \cr
\noalign{\vskip 5pt}%
g_m\cr} = 0 \ \ .\cr}\eqno(4.3)$$
\noindent From (4.3) the regular solution of $f_m$ in the inside region is
$$f^{\rm in}_m (r) = J_{|m|} (k_0 r) \eqno(4.4)$$
where
$$k_0 = \left( E - {\xi \over R}\right)^2 - M^2 \quad . \eqno(4.5)$$
In the outside region Eq. (4.3) can be written as
$$\eqalign{\bigg[ {d^2 \over dr^2} + {1 \over r}\ {d \over dr} &- {\eta +
(m+\alpha)^2 \over r^2} + \bigg( E - {\xi \over r}\bigg)^2\cr
&- M^2 + {1 \over r^2}\ (\eta \sigma_3 - i \xi \sigma_1 ) \bigg]
\pmatrix{f_m\cr
\noalign{\vskip 5pt}%
g_m\cr} = 0 \quad .\cr}\eqno(4.6)$$
Again one diagonalizes the $1/r^2$ term using the result
$$\eta \sigma_3 - i \xi \sigma_1 = \epsilon (\eta) \sqrt{\eta^2 -
\xi^2}\ U^{-1} \sigma_3 U \eqno(4.7)$$
where
$$\eqalign{U^{\pm 1} &= a_+ \mp \epsilon (\eta) a_- \sigma_2 \cr
\noalign{\vskip 4pt}%
a_\pm &= \pm {1 \over \sqrt{2}} \ \bigg[ {|\eta | \over \sqrt{\eta^2 -
\xi^2}} \pm 1 \bigg]^{{1 \over 2}} \quad .\cr}\eqno(4.8)$$
By defining
$$\chi_m \equiv \pmatrix{ \chi_{_{3m}} \cr
\noalign{\vskip 5pt}%
\chi_{_{4m}}\cr} = U \pmatrix{f_m \cr
\noalign{\vskip 5pt}%
g_m\cr}$$
the second-order equations
 for $\chi_{_{3m}}$ and $\chi_{_{4m}}$ can be solved
to yield
$$\eqalign{\chi_{_{3m}} = \ &A_m e^{ikr} (-2ikr)^{c_m} F \bigg( c_m + {1 \over
2} + {iE\xi \over k}\  | 2 c_m + 1 | -2ikr\bigg)\cr
\noalign{\vskip 4pt}%
& + B_m e^{ikr} (-2ikr)^{-c_m} F \bigg( -c_m + {1 \over
2} + {iE\xi \over k}\  | -2 c_m  | -2ikr\bigg)\cr
\noalign{\vskip 4pt}%
\chi_{_{4m}} =\  &C_m e^{ikr} (-2ikr)^{d_m} F \bigg( d_m + {1 \over
2} + {iE\xi \over k}\  | 2 d_m + 1 | -2ikr\bigg)\cr
\noalign{\vskip 4pt}%
& + D_m e^{ikr} (-2ikr)^{-d_m} F \bigg( -d_m + {1 \over
2} + {iE\xi \over k}\  | 1-2 d_m  | -2ikr\bigg)\cr}\eqno(4.9)$$
where
$$\eqalign{c_m &= \sqrt{\eta^2 - \xi^2} - {\epsilon (\eta) \over 2}\cr
\noalign{\vskip 4pt}%
d_m &= \sqrt{\eta^2 - \xi^2} + {\epsilon (\eta) \over
2} \quad .\cr}\eqno(4.10)$$
Upon obtaining the relations between coefficients
$$\eqalign{C_m &= \Omega_3 A_m\cr
D_m &= \Omega_4 B_m\cr}\eqno(4.11)$$
for $\epsilon (\eta ) = 1$, and
$$\eqalign{A_m &= \Omega_3 C_m\cr
B_m &= \Omega_4 D_m\cr}\eqno(4.12)$$
for $\epsilon (\eta ) = -1$
where
$$\eqalign{\Omega_3 &= {-k \big( \eta^2 - \xi^2 + E^2 \xi^2
k^{-2}\big)
 \over
2(\eta^2 - \xi^2) (2 \sqrt{\eta^2 - \xi^2} +1)} \ \bigg( E+ {\eta \over
\sqrt{\eta^2 - \xi^2}} \ M \bigg)^{-1}\cr
\noalign{\vskip 4pt}%
\Omega_4 &= 2k \big( 2 \sqrt{\eta^2 - \xi^2} - 1 \big)
\bigg(E+ {\eta \over \sqrt{\eta^2 - \xi^2}} \ M \bigg)^{-1}\quad
,\cr}\eqno(4.13)$$
one can proceed to match the boundary conditions at $r=R$.  This yields
once again the result
$$B_m = D_m = 0 $$
in the absence of a fine tuning condition on $\xi$.  Thus the radial wave
equation becomes
$$\eqalign{f_m(r) =
A_m e^{ikr} \bigg[&a_+
(-2ikr)^{\gamma_{m,s}} F \bigg( \gamma_{m,s} + {1 \over 2} +
{iM\xi \over k} | 2\gamma_{m,s} + 1 | -2ikr\bigg)\cr
&-a_-  \Omega_3 (-2i kr)^{\gamma_{m,s}+1}
F\bigg( \gamma_{m,s} + {3 \over 2} + {iM\xi \over k} |
2 \gamma_{m,s} + 3 | -2ikr \bigg)\bigg]\cr}\eqno(4.14)$$
for $\epsilon (\eta ) =1$, and
$$\eqalign{f_m (r) =
C_m e^{ikr} \bigg[ &a_+ \Omega_3
(-2ikr)^{\gamma_{m,s}+1} F \bigg( \gamma_{m,s} + {3 \over 2} +
{iM\xi \over k} | 2\gamma_{m,s} + 3 | -2ikr\bigg)\cr
&+ ia_- (-2i kr)^{\gamma_{m,s}}
F\bigg( \gamma_{m,s} + {1 \over 2} + {iM\xi \over k} |
2 \gamma_{m,s} + 1 | -2ikr \bigg)\bigg]\cr}\eqno(4.15)$$
 for $\epsilon (\eta ) = -1$ where
$$\gamma_{m,s} = \sqrt{\eta^2 - \xi^2} - {1 \over 2} \quad .\eqno(4.16)$$
In the case of the
discontinuous Coulomb potential the radial wave function is
 once again given by Eqs. (4.14)
 and (4.15) in close analogy to the  scalar coupling theory.
 In each case the bound state energies  are found to be
$$E^2_V = M^2 \left[ 1 +
 {\xi^2 \over \left[ n-1 + \sqrt{\left[ s (m +\alpha) +
{1 \over 2}\right]^2 - \xi^2}\right]^2} \right]^{-1}
\qquad n = 1, 2, 3, \dots\eqno(4.17)$$

\noindent This spectrum is in agreement with the result obtained
 in ref. 10 without reference to the vanishing radius flux tube method.

It is of interest to note that the result (2.39) for the scalar binding
energies $E_S$ and (4.17) can be combined in the single expression
$$E_{S \choose V} = M \left\{ 1 \mp {\xi^2 \over \left[ n-1 +
\left\{\left[s \left( m + \alpha \right) + {1 \over 2} \right]^2
\pm \xi^2 \right\}^{1 \over 2}\right]^2} \right\}^{\pm {1 \over 2}}
\quad .\eqno(4.18)$$
This compact form allows one to discuss simultaneously the significance
of the results obtained in  the
 various coupling models considered
here.
\medskip

\noindent {\bf 5. Conclusion}

The method of the finite radius flux tube was advanced
 originally as a possible tool
to deal with complications encountered in the relativistic AB calculation
for spin-1/2 particles.  It was successful in that application as well as
in the corresponding Galilean ABC problem.  In particular it was found [7]
in the latter case that when the Coulomb potential is attractive bound
state energies occur at
$${\cal E}_n = - {1 \over 2}\ {M \xi^2 \over (n - {1 \over 2}\ \pm
|m +\alpha |)^2} \qquad n = 1,2, \dots \eqno(5.1)$$
where the minus sign is realized only in the case $|m + \alpha | < {1
\over 2}$, $\alpha s < 0$.  The corresponding result for the ABCD problem
(i.e., the Dirac equation treatment of the ABC potential) is given by
(4.18) and there arises the question as to the mutual consistency of these
expressions.

Comparison of the two formulae in the Galilean limit indicates that
compatibility requires that the form $|m + \alpha + {s \over 2} |-1$ of the
ABCD case be equivalent to $\pm |m + \alpha |-{1 \over 2}$ of the ABC
result.  These are not obviously the same and in fact they cannot be
generally equivalent.  The expression (4.18) has the property that it is
invariant under the replacement
$$|m + \alpha | \rightarrow -|m + \alpha | - s \epsilon (m + \alpha)$$
whereas (5.1) does not.  More simply, for $\alpha =0$ the nonrelativistic
energies (5.1) become spin independent unlike the relativistic ones.  More
detailed analysis of the ABCD spectrum shows that even though there are
singular states in the general case (corresponding to the minus sign choice
in (5.1)), the details of the spectrum do not agree in the limit in which $c
$ (the velocity of light) becomes arbitrarily large.

It is not difficult to trace the source of this discrepancy.  As has been
emphasized during the course of the calculations presented here, the very
different qualitative behaviors of the Galilean and the relativistic wave
functions for $\xi \not= 0$ mean that the $R \rightarrow 0$ and $c
\rightarrow \infty$ limits do not commute.  Thus the singular solutions
which arise in the Galilean limit from the delta function magnetic field
 are in the ABCD calculation a result of purely relativistic effects in the
$\xi \not= 0$ case.  In fact the magnetic field
term had no effect whatever upon the structure of the relativistic wave
functions in the $R=0$ limit.

There is, of course, no {\it a priori} requirement that a relativistic wave
equation give totally satisfactory results in describing a physical
phenomenon which presumably demands a field theoretical approach for full
consistency.  The strains put upon wave mechanics by such phenomena as
Klein's paradox are, for example, well known.  On the other hand one
generally expects relativistic wave equations to give reliable results in
the Galilean limit.  That has not happened in the present case for reasons
which have been carefully documented.  Whether the shortcoming is in the
finite radius flux tube method or in the nature of relativistic wave
equations -- or even whether it makes sense to attempt to separate these
two issues -- is not obvious.  This paper does, however, clearly show that
one is now pressing hard upon the limits of joint applicability of these
two calculational techniques.

\medskip

\noindent {\bf Acknowledgment}

This work is supported in part by the U.S. Department of Energy Grant No.
DE-FG-02-91ER40685. D.K.P. would like to thank the Korean Science and
Engineering Foundation.

\vfill\eject
\medskip

\noindent {\bf References}

\medskip

	\item{1.} Y. Aharonov and D. Bohm, Phys. Rev. {\bf 115}, 485 (1959).

	\item{2.} M. Peshkin and A. Tonomura, The Aharonov-Bohm Effect,
	(Springer-Verlag, Berlin, 1989).

	\item{3.} F. Wilczek, Phys. Rev. Lett. {\bf 48}, 1144 (1982).

	\item{4.} Ph. de Sousa Gerbert, Phys. Rev. {\bf D40}, 1346 (1989).

	\item{5.} C.R. Hagen, Phys. Rev. Lett. {\bf 64}, 503 (1990), Int. J.
	Mod. Phys. {\bf A6}, 3119 (1991).

	\item{6.} F.A.B. Coutinho and J.F. Perez, Phys. Rev. {\bf D48}, 932
	(1993).

	\item{7.} C.R. Hagen, Phys. Rev. {\bf D48}, 5935 (1993).

	\item{8.} J.M. L\'evy-Leblond, Commun. Math. Phys. {\bf 6}, 286 (1967).

	\item{9.} D.K. Park, HEP-TH9405009 (1994).

	\item{10.} S.H. Guo, X.L. Yang, F.T. Chan, K.W. Wong, and W.Y.
         Ching,
	Phys. Rev.  {\bf A43}, 1197  (1991).

\vfill\eject

\baselineskip=18pt

\midinsert
$$\vbox{\offinterlineskip
\halign{&\vrule#&\strut\ #\ \cr
\multispan{6}\hfil TABLE I. Leading terms in $u_1 (R)$.\hfil\cr
\noalign{\bigskip}
\noalign{\hrule}
height7pt&\omit&&\omit&&\omit&\cr
&\hfil $u_1(R)(\epsilon(\eta) =1)$\hfil&&\hfil First leading term\hfil
&&\hfil Second
leading term\hfil&\cr
height7pt&\omit&&\omit&&\omit&\cr
\noalign{\hrule}
height7pt&\omit&&\omit&&\omit&\cr
&Relativistic ABC \hfil
 &&$\cos {\theta \over 2}\ (-2ikR)^{\beta_{m,s}}$\hfil
&&$(-2ikR)^{\beta_{m,s}+1} \left[\matrix{{(iM \xi/k) \over 1+ 2\beta_{m,s}}\
\cos {\theta \over 2}\cr
\noalign{\vskip 4pt}%
+i \sin {\theta \over 2}\ \Omega_1\cr}\right]$&\cr
height7pt&\omit&&\omit&&\omit&\cr
\noalign{\hrule}
height7pt&\omit&&\omit&&\omit&\cr
&Relativistic AB \hfil &&$(-2ikR)^{s(m+\alpha)}$ \hfil
&&$O \left(R^{s(m+\alpha)+2}\right)$&\cr
height7pt&\omit&&\omit&&\omit&\cr
\noalign{\hrule}
height7pt&\omit&&\omit&&\omit&\cr
&Galilean ABC \hfil &&$(-2ikR)^{s(m+\alpha)}$ \hfil
&&${iM\xi/k \over 1+2s(m+\alpha)} \
(-2ikR)^{s(m+\alpha)+1}$&\cr
height7pt&\omit&&\omit&&\omit&\cr
\noalign{\hrule}\noalign{\medskip}
\hfil\cr}}$$
\endinsert

\vskip 1 truein

\midinsert
$$\vbox{\offinterlineskip
\halign{&\vrule#&\strut\ #\ \cr
\multispan{6}\hfil TABLE II.  Leading terms in $u_2 (R)$.\hfil\cr
\noalign{\bigskip}
\noalign{\hrule}
height7pt&\omit&&\omit&&\omit&\cr
&\hfil $u_2(R)(\epsilon(\eta) =1)$\hfil&&\hfil First leading term\hfil
&&\hfil Second
leading term\hfil&\cr
height7pt&\omit&&\omit&&\omit&\cr
\noalign{\hrule}
height7pt&\omit&&\omit&&\omit&\cr
&Relativistic ABC \hfil
 &&$i \sin {\theta \over 2}\ \Omega_2 (-2ikR)^{-\beta_{m,s} -1}$\hfil
&&$(-2ikR)^{-\beta_{m,s}} \left[\matrix{\cos {\theta \over 2}\cr
\noalign{\vskip 4pt}%
+ {M \xi /k  \over 1+2 \beta_{m,s}}\sin {\theta \over
2}\ \Omega_2\cr}\right]$&\cr
height7pt&\omit&&\omit&&\omit&\cr
\noalign{\hrule}
height7pt&\omit&&\omit&&\omit&\cr
&Relativistic AB \hfil &&$(-2ikR)^{-s(m+\alpha)}$ \hfil
&&$O \left(R^{-s(m+\alpha)+2}\right)$&\cr
height7pt&\omit&&\omit&&\omit&\cr
\noalign{\hrule}
height7pt&\omit&&\omit&&\omit&\cr
&Galilean ABC \hfil &&$(-2ikR)^{-s(m+\alpha)}$ \hfil
&&${iM\xi/k \over 1-2s(m+\alpha)} \
(-2ikR)^{-s(m+\alpha)+1}$&\cr
height7pt&\omit&&\omit&&\omit&\cr
\noalign{\hrule}\noalign{\medskip}
\hfil\cr}}$$
\endinsert

\vfill\eject

\midinsert
$$\vbox{\offinterlineskip
\halign{&\vrule#&\strut\ #\ \cr
\multispan{6}\hfil TABLE III. Leading terms in $v_1(R)$, [$\Omega^G_1$ is
Galilean limit of $\Omega_1$].\hfil\cr
\noalign{\bigskip}
\noalign{\hrule}
height7pt&\omit&&\omit&&\omit&\cr
&\hfil $v_1(R)(\epsilon(\eta) =-1)$\hfil&&\hfil First leading term\hfil
&&\hfil Second
leading term\hfil&\cr
height7pt&\omit&&\omit&&\omit&\cr
\noalign{\hrule}
height7pt&\omit&&\omit&&\omit&\cr
&Relativistic ABC \hfil
 &&$-i \sin {\theta \over 2}\  (-2ikR)^{\beta_{m,s}}$\hfil
&&$(-2ikR)^{\beta_{m,s}+1} \left[\matrix{\cos {\theta \over 2}\ \Omega_1\cr
\noalign{\vskip 4pt}%
+ {M \xi /k  \over 1+2 \beta_{m,s}}\sin {\theta \over
2}\cr}\right]$&\cr
height7pt&\omit&&\omit&&\omit&\cr
\noalign{\hrule}
height7pt&\omit&&\omit&&\omit&\cr
&Relativistic AB \hfil &&$\Omega_1 (\xi =0)(-2ikR)^{-s(m+\alpha)}$ \hfil
&&$O \left(R^{-s(m+\alpha)+2}\right)$&\cr
height7pt&\omit&&\omit&&\omit&\cr
\noalign{\hrule}
height7pt&\omit&&\omit&&\omit&\cr
&Galilean ABC \hfil &&$\Omega^G_1 (-2ikR)^{-s(m+\alpha)}$ \hfil
&&$\Omega^G_1 (-2ikR)^{-s(m+\alpha)+1} \ {iM\xi /k \over
1-2 s(m+\alpha)}$&\cr
height7pt&\omit&&\omit&&\omit&\cr
\noalign{\hrule}\noalign{\medskip}
\hfil\cr}}$$
\endinsert

\smallskip

\vskip 1 truein

\midinsert
$$\vbox{\offinterlineskip
\halign{&\vrule#&\strut\ #\ \cr
\multispan{6}\hfil TABLE IV. Leading terms in $v_2 (R)$,
[$\Omega_2^G$ is Galilean limit of $\Omega_2$].\hfil\cr
\noalign{\bigskip}
\noalign{\hrule}
height7pt&\omit&&\omit&&\omit&\cr
&\hfil $v_2(R)(\epsilon(\eta) =-1)$\hfil&&\hfil First leading term\hfil
&&\hfil Second
leading term\hfil&\cr
height7pt&\omit&&\omit&&\omit&\cr
\noalign{\hrule}
height7pt&\omit&&\omit&&\omit&\cr
&Relativistic ABC \hfil
 &&$\cos {\theta \over 2}\ \Omega_2 (-2ikR)^{-\beta_{m,s}-1}$\hfil
&&$-(-2ikR)^{-\beta_{m,s}} \left[\matrix{{iM \xi/k \over 1+ 2\beta_{m,s}}\
\cos {\theta \over 2}\ \Omega_2\cr
\noalign{\vskip 4pt}%
+i \sin {\theta \over 2}\cr}\right]$&\cr
height7pt&\omit&&\omit&&\omit&\cr
\noalign{\hrule}
height7pt&\omit&&\omit&&\omit&\cr
&Relativistic AB \hfil &&$\Omega_2 (\xi =0)(-2ikR)^{s(m+\alpha)}$ \hfil
&&$O \left(R^{s(m+\alpha)+2}\right)$&\cr
height7pt&\omit&&\omit&&\omit&\cr
\noalign{\hrule}
height7pt&\omit&&\omit&&\omit&\cr
&Galilean ABC \hfil &&$\Omega^G_2(-2ikR)^{s(m+\alpha)}$ \hfil
&&$(-2ikR)^{s(m+\alpha)+1} \ {iM\xi /k \over
1+2 s(m+\alpha)} \ \Omega_2^G$&\cr
height7pt&\omit&&\omit&&\omit&\cr
\noalign{\hrule}\noalign{\medskip}
\hfil\cr}}$$
\endinsert

\smallskip

\vfill\eject

\midinsert
$$\vbox{\offinterlineskip
\halign{&\vrule#&\strut\ #\ \cr
\multispan{6}\hfil TABLE V. Leading terms in $J_{|m|} (k_0 R)$
$\left[ k_0^G = \left[ 2m \left( E - {\xi \over R}\right) \right]^{1/2}
\right]$.\hfil\cr
\noalign{\bigskip}
\noalign{\hrule}
height7pt&\omit&&\omit&&\omit&\cr
&\hfil $J_{|m|} (k_0 R)$\hfil&&\hfil First leading term\hfil
&&\hfil Second
leading term\hfil&\cr
height7pt&\omit&&\omit&&\omit&\cr
\noalign{\hrule}
height7pt&\omit&&\omit&&\omit&\cr
&Relativistic ABC \hfil
 &&$i^{|m|}I_{|m|} (\xi)$\hfil
&&$ i^{|m|} M I^\prime_{|m|} (\xi)R$&\cr
height7pt&\omit&&\omit&&\omit&\cr
\noalign{\hrule}
height7pt&\omit&&\omit&&\omit&\cr
&Relativistic AB \hfil &&${\left( {\sqrt{E^2 -M^2} \over 2}\right)^{|m|} \over
\Gamma (1+ |m|)}\ R^{|m|}$ \hfil
&&$- {\left( {\sqrt{E^2 -M^2} \over 2}\right)^{|m| +2} \over
\Gamma (2 +|m|)}\ R^{|m| +2}$&\cr
height7pt&\omit&&\omit&&\omit&\cr
\noalign{\hrule}
height7pt&\omit&&\omit&&\omit&\cr
&Galilean ABC \hfil &&${(k^G_0 R)^{|m|} \over 2^{|m|} \Gamma (1+
|m|)}$ \hfil
&&$- {(k^G_0 R)^{|m| +2} \over 2^{|m| +2} \Gamma (2+ |m|)}$&\cr
height7pt&\omit&&\omit&&\omit&\cr
\noalign{\hrule}\noalign{\medskip}
\hfil\cr}}$$
\endinsert

\smallskip

\vskip 1.5 truein

\midinsert
$$\vbox{\offinterlineskip
\halign{&\vrule#&\strut\ #\ \cr
\multispan{6}\hfil TABLE VI. Leading terms in $\left( {d \over dR} +
 {\alpha s \over R}\right)$ $J_{|m|} (k_0 R)$.\hfil\cr
\noalign{\bigskip}
\noalign{\hrule}
height7pt&\omit&&\omit&&\omit&\cr
&\hfil $\left({d \over dR} + {\alpha s \over R}\right) J_{|m|}
(k_0 R)$ \hfil&&\hfil First leading term\hfil
&&\hfil Second
leading term\hfil&\cr
height7pt&\omit&&\omit&&\omit&\cr
\noalign{\hrule}
height7pt&\omit&&\omit&&\omit&\cr
&Relativistic ABC \hfil
 &&$i^{|m|} \ {(|m| + \alpha s) I_{|m|} (\xi) + \xi I_{|m|+1}
(\xi) \over R}$\hfil
&&$ i^{|m|} M \pmatrix{(|m| + \alpha s) I^\prime_{|m|} (\xi)\cr
\noalign{\vskip 4pt}%
+ I_{|m|+1} (\xi)\cr
\noalign{\vskip 4pt}%
+ \xi I^\prime_{|m|+1} (\xi)\cr}$&\cr
height7pt&\omit&&\omit&&\omit&\cr
\noalign{\hrule}
height7pt&\omit&&\omit&&\omit&\cr
&Relativistic AB \hfil &&$(|m| + \alpha s)\ {\left( {\sqrt{E^2-M^2}\over 2}
\right)^{|m|} \over \Gamma (1+ |m|)} \ R^{|m|-1}$ \hfil
&&$O \left( R^{|m| +1}\right)$&\cr
height7pt&\omit&&\omit&&\omit&\cr
\noalign{\hrule}
height7pt&\omit&&\omit&&\omit&\cr
&Galilean ABC \hfil &&$(|m| +\alpha s)\ {(k^G_0)^{|m|} \over 2^{|m|}
\Gamma (1+ |m|)}\ R^{|m| -1}$ \hfil
&&$O \left( (k^G_0 )^{|m| +2} R^{|m|+1}\right)$&\cr
height7pt&\omit&&\omit&&\omit&\cr
\noalign{\hrule}\noalign{\medskip}
\hfil\cr}}$$
\endinsert

\vfill\eject

\midinsert
$$\vbox{\offinterlineskip
\halign{&\vrule#&\strut\ #\ \cr
\multispan{6}\hfil TABLE VII.  Leading terms in $u (R)$.\hfil\cr
\noalign{\bigskip}
\noalign{\hrule}
height7pt&\omit&&\omit&&\omit&\cr
&\hfil $u(R)$\hfil&&\hfil First leading term\hfil
&&\hfil Second
leading term\hfil&\cr
height7pt&\omit&&\omit&&\omit&\cr
\noalign{\hrule}
height7pt&\omit&&\omit&&\omit&\cr
&Relativistic ABC \hfil&&{}&&{}&\cr
&$\ \ \ \ (ms \geq 0)$\hfil
 &&${k^{|m|} R^{|m|-1} \over 2^{|m|} \Gamma (1+ |m|)} \ (|m|+\alpha s)$\hfil
&&$\xi \sqrt{{E-M \over E+M}}\  {k^{|m|+1} R^{|m|} \over
2^{|m|+1} \Gamma (2+|m|)}$&\cr
height7pt&\omit&&\omit&&\omit&\cr
\noalign{\hrule}
height7pt&\omit&&\omit&&\omit&\cr
&Relativistic ABC \hfil&&{}&&{}&\cr
&$\ \ \ \ (ms < 0)$ \hfil&&$- \xi \sqrt{{E-M \over E+M}}
{k^{|m|-1} R^{|m|-2} \over 2^{|m|-1} \Gamma (|m|)}$\hfil
&&${k^{|m|} R^{|m|-1} \over
2^{|m|} \Gamma (1+ |m|)}\  (|m| + \alpha s)$&\cr
height7pt&\omit&&\omit&&\omit&\cr
\noalign{\hrule}
height7pt&\omit&&\omit&&\omit&\cr
&Relativistic AB \hfil&&${k^{|m|} R^{|m|-1} \over 2^{|m|} \Gamma
(1+|m|)} \ (|m|+\alpha s)$\hfil
&&$O \left( R^{|m|+1}\right)$&\cr
height7pt&\omit&&\omit&&\omit&\cr
\noalign{\hrule}
height7pt&\omit&&\omit&&\omit&\cr
&Galilean ABC \hfil &&${k^{|m|}R^{|m|-1} \over 2^{|m|} \Gamma
(1+|m|)}\ (|m|+\alpha s)$\hfil
&&$O \left( R^{|m|+1}\right)$&\cr
height7pt&\omit&&\omit&&\omit&\cr
\noalign{\hrule}\noalign{\medskip}
\hfil\cr}}$$
\endinsert

\end